\documentclass[pra,twocolumn]{revtex4}

\usepackage[title]{appendix}
\usepackage{graphicx}
\usepackage{amsfonts}
\usepackage{textcomp}
\usepackage{float}
\usepackage{amsmath}
\usepackage[utf8]{inputenc}
\newcommand{\ket}[1]{\left|#1\right>}
\newcommand{\bra}[1]{\left< #1 \right|}
\newcommand{\beq}{\begin{equation}}
\newcommand{\eeq}{\end{equation}}
\newcommand{\HH}{\hat{H}}
\newcommand{\mean}[1]{\langle{#1}\rangle{}}

\begin{document}

\title{Readout of a dopant spin in the anisotropic quantum dot with a single magnetic ion}

\author{A. Rodek}
\email{Aleksander.Rodek@fuw.edu.pl}
\author{T. Kazimierczuk}
\email{Tomasz.Kazimierczuk@fuw.edu.pl}
\author{A. Bogucki}
\author{T. Smoleński}
\author{W. Pacuski}
\author{P. Kossacki}

\address{Institute of Experimental Physics, Faculty of Physics,
University of Warsaw, ul. Pasteura 5, 02-093 Warsaw, Poland}

\date{\today}

\begin{abstract}
Owing to exchange interaction between the exciton and magnetic ion, a quantum dot embedding a single magnetic ion is a great platform for optical control of individual spin. In particular, a~quantum dot provides strong and sharp optical transitions, which give experimental access to spin states of an individual magnetic ion. We show, however, that physics of quantum dot excitons also complicate spin readout and optical spin manipulation in such a~system. This is due to electron-hole exchange interaction in anisotropic quantum dots, which affects the polarization of the emission lines. One of the consequences is that the intensity of spectral lines in a single spectrum are not simply proportional to the population of various spin states of magnetic ion. In order to provide a solution of the above problem, we present a method of extracting both the spin polarisation degree of a neutral exciton and magnetic dopant inside a semiconductor quantum dot in an external magnetic field. Our approach is experimentally verified on a system of CdSe/ZnSe quantum dot containing a single Fe$^{2+}$ ion. Both the resonant and non-resonant excitation regimes are explored resulting in a record high optical orientation efficiency of dopant spin in the former case. The proposed solutions can be easily expanded to any other system of quantum dots containing magnetic dopants.
\end{abstract}

\maketitle

\section{Introduction}
Spin degree of freedom of defect centers in semiconductors is desirable storage for quantum information thanks to relatively long coherence times caused by weak interaction with the crystal environment \cite{DOHERTY20131,Flatte_NM11,Awschalom_S13}.
Particular example of such defects are transition metal (TM) dopants, which feature non-zero spin due to partially filled electron \textit{d} shell. Extensive studies of semiconductors doped with TM ions (diluted magnetic semiconductors)
established a firm relation between the energy of the excitonic transition and the average spin of the TM ions (giant Zeeman effect \cite{gaj}). Because of a large number of involved dopants in bulk crystals, the exciton energy shift (and thus the average TM-dopant spin state) could be considered a continuous variable.
This situation has changed after the introduction of a system of a single TM ion in a quantum dot (QD), in which the spin quantization was directly reflected by the quantization of localized exciton energy\cite{besombes}.
The photoluminescence (PL) spectrum of the exciton in such a case consists of a series of lines, which, due to the axial character of the heavy holes in the QDs, can be mapped to given
projections of the TM ion onto the growth axis\cite{besombes, kudelski, kobak2014,Lafuente_PRB2016}.
Consequently, relative intensity of each of these lines constitutes a convenient measure of the orientation of the TM ion spin.
This particular feature has been exploited in a number of studies concerning various types of dopants in different semiconductor systems \cite{Gall_PRL2009,Goryca_PRL2009,kobak2014,smolen2015,gorycaprb,smolennewsystem,Smolenski2016,kacper,smolen2018,besombes_prb_2008}.
Such an approach is naturally limited to the case of the non-resonant excitation, since in the resonant case the directly excited state is vastly over-represented
in the photoluminescence spectrum\cite{besombesresonant, besombes_prb_2017}.

In this work we demonstrate that even for the non-resonant excitation it is not sufficient to use the relative intensities of the emission lines
as a measure of the spin orientation of the TM ion embedded in a QD. As we show, the simple picture is invalidated by the interplay between
optical orientation of the exciton and the non-trivial dependence of the TM ion spin on the selection rules in anisotropic quantum dots.  Our considerations will be showcased using a particular system of a CdSe/ZnSe QD with single Fe$^{2+}$ ion \cite{Smolenski2016}, but the same reasoning should be valid also for all other known systems of that kind \cite{krebsprb_2016,kobakprb_2018,besombesprb_2019}.

\begin{figure}
\includegraphics[scale=1]{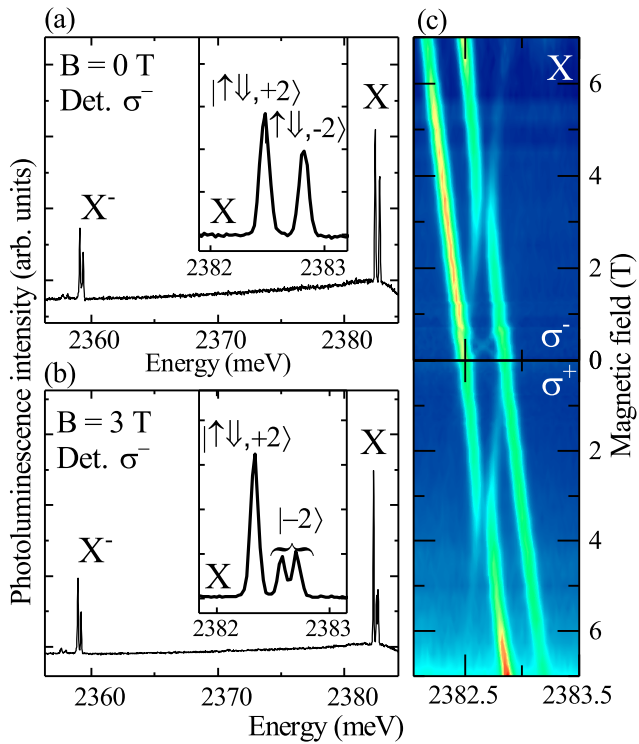}
\caption{(a-b) $\sigma^{-}-$ polarised photoluminescence of a quantum dot with a single Fe$^{2+}$ ion in a magnetic field of (a) 0 and (b) 3 T. Inset shows the close-up of the neutral exciton lines.
(c) Neutral exciton intensity as a function of magnetic field from 0 to 6 T that shows characteristic anticrossing between exciton states of opposite spins: $\downarrow\Uparrow$ and $\uparrow\Downarrow$. \label{fig1}}
\end{figure}

\section{Experiment}

The illustratory experimental data used in this work was measured in a photoluminescence (PL) setup with high spatial resolution obtained using Cassagrain-type microscope
with effective $\mathrm{NA}=0.7$ \cite{SEPIOL1997444}. The sample was cooled down to $T=1.8$~K inside a superconductive magnet, which provided magnetic field up to
$B=10$~T in Faraday configuration. The sample was excited quasi-resonantly at $E=2407.44$~meV.  The PL signal was detected in two circular polarizations using a motorized $\lambda/4$ waveplate and a linear polarizer. The sample structure with CdSe: Fe QDs embedded in ZnSe barrier was described in Ref. \onlinecite{Smolenski2016}.

The characteristics of measured PL spectra were consistent with previous studies \cite{Smolenski2016, bayerfss}. The spectrum of a single non-resonantly
excited QD with a single Fe$^{2+}$ ion shown in Fig. \ref{fig1}(a) features three pairs of lines. Each pair corresponds to recombination of a certain excitonic
complex: neutral exciton (X), charged exciton (X$^-$), or neutral biexciton (XX). In the following analysis we focus on the neutral exciton as the fundamental
transition in the neutral QD.

\section{Paradox of simplified determination of the ion orientation}

The splitting of the neutral exciton line in QDs with single TM ions is a reflection of the exciton-ion exchange interaction. In the analyzed case, the lower (higher) energy
line corresponds to state with anti-parallel (parallel) orientation of exciton and Fe$^{2+}$ spins. Due to optical selection rules, by detecting $\sigma-$ circularly polarized light, we study excitons build of a spin-up electron ($\uparrow$) and a spin-down heavy hole ($\Downarrow$). As a consequence, we can unequivocally attribute the lower (higher) energy line in $\sigma-$ polarization to the particular spin projection of the iron ion: $S=+2$ ($S=-2$), as marked in insets of Fig.~\ref{fig1}a. We note that this link is valid even if the eigenstates of the system does not correspond to the pure spin states. It is particularly evident, in the magnetic field of a few Tesla in Faraday configuration, near the anticrossing caused by the anisotropic interaction between the carriers (see Fig.~\ref{fig1}b). In such a field anti-crossed lines are linearly-polarized, but still correspond to well-defined projection of the iron spin.

The described correspondence is a foundation of a go-to method of measuring the orientation of the dopant spin $\eta_\mathrm{Fe}$ simply by taking a normalised difference in the intensities of lower and higher energy emission peaks \cite{papierska1,pilat1,gorycaprb}:
\begin{equation}
\eta_\mathrm{Fe} = 2\frac{I_{S=+2}-I_{S=-2}}{I_{S=+2}+I_{S=-2}}
\label{eq:stopien}
\end{equation}
where $I_{S=\pm 2}$ denotes the intensity of the line(s) related to a $S=\pm 2$ spin of the Fe$^{2+}$ ion in a given circular polarisation that defines the spin of exciton.
Such a method was applied to the case of neutral exciton \cite{papierska1, gorycaprb, smolen2015}, charged exciton \cite{pilat1}, and biexciton transition\cite{smolen2015}. It was identified that the last case is the most reliable, as the biexciton due to its singlet nature it is not affected by the issue of exciton spin relaxation\cite{smolen2015}.
Unfortunately, the biexciton transition is not always available, e.g. due to overlap with other features in the spectrum or simply due to insufficient intensity.

\begin{figure}[H]
\includegraphics[width=0.5\textwidth]{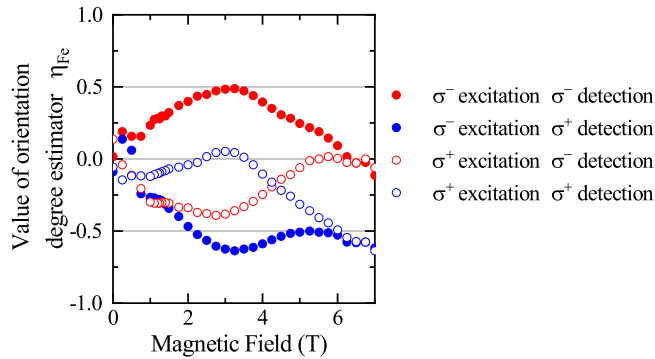}
\label{fig2}
\caption{Orientation of the iron spin estimated using applying the formula \ref{eq:stopien} to data measured using different combinations of circular polarization of excitation and detection. Please note the disparity between datasets corresponding to the same excitation condition.
}

\end{figure}

For that reason the described method of extracting Fe$^{2+}$ spin orientation was applied for the quantum dot presented in Fig. \ref{fig1} using the neutral exciton transition. The expression \ref{eq:stopien} was applied independently in all four combinations of circular polarization of excitation and detection. The resulting values of this procedure (i.e., apparent ion spin orientation degree) are presented in Fig. \ref{fig2}. As clearly seen, the orientation degree $\eta$ extracted from various combinations of polarization yields significantly different values. As discussed earlier, using the neutral exciton instead of the biexciton in this procedure introduces a bias due to spin relaxation during exciton lifetime. However, such a bias should not depend on the excitation conditions, therefore it cannot explain the difference between $\eta$ values obtained for $\sigma^+/\sigma^-$ and $\sigma^-/\sigma^-$ polarizations (using notation of "polarization of excitation / polarization of detection").
Na\"ive interpretation would attribute this difference to the Fe$^{2+}$-spin optical orientation effect, i.e., to dependence of the direction of the dopant spin on the helicity of the excitation. In this frame the collected data could lead to conclusion that the maximum efficiency of the optical orientation (largest difference between full and empty red symbols in Fig. \ref{fig2}) occurs in the field of $B\approx 3$~T. Yet, such interpretation is ostensibly contradicted by data in the $\sigma^+/\sigma^+$ and $\sigma^-/\sigma^+$ polarizations, which seemingly indicates \emph{opposite} optical orientation.

Described disparity can lead to counter-intuitive effect in the experiment, e.g., while probing optical Fe$^{2+}$ orientation effect at $B=3$ T using $\sigma^-$ polarization. In such a~setting ratio of intensity of PL signal related to $\ket{2+}$ spin state to the intensity of PL signal related to $\ket{2-}$ spin state \emph{decreases} when switching from $\sigma^-$ to $\sigma^+$ excitation, which can be easily misinterpreted as a proof of negative optical pumping of the ion spin.

The explanation of this paradox is directly related to \emph{exciton} spin optical orientation. In particular, the relative intensities of the studied lines in the PL spectrum depend not only on the Fe$^{2+}$ spin, but also on how well the polarization of the involved exciton state agrees with the polarization of the excitation. While at zero or at high magnetic field both emission lines exhibit the same polarization degree, in the intermediate range one of the lines is affected by the anticrossing due to electron-hole exchange interaction \cite{Smolenski2016}. For the presented QD this anticrossing occurs about $B\approx 3$~T, which is clearly seen in Fig. \ref{fig1}(c).

In the next section we describe the theoretical model of this phenomenon and propose a procedure to extract the real orientation degree of the dopant spin from the collected data.

\section{Proposed model \label{sec:model}}
Presented method of calculating the effective degree of iron ion and exciton spin polarisation is based on the hamiltonian introduced in Refs. \onlinecite{Smolenski2016, smolenskiprb_2017}. In this approach we study a hamiltonian describing an interacting pair of bright exciton and an~Fe$^{2+}$ ion of spin $\mean{S_z}=~\pm 2$ in a magnetic field of Faraday configuration. For a four-dimensional basis of different spin configurations $\ket{S_\mathrm{X},S_\mathrm{Fe}}=\{\ket{\downarrow\Uparrow,+2},\ket{\downarrow\Uparrow,-2},
\ket{\uparrow\Downarrow,+2},\ket{\uparrow\Downarrow,-2}\}$ it takes on the following form:
\begin{equation}
\HH=
\frac{1}{2}
\begin{bmatrix}
\Delta_{s-p,d}   & \delta_1 & \delta_\mathrm{Fe} & 0 \\
 \delta_1 & -\Delta_{s-p,d}  & 0 & \delta_{1}\\
\delta_\mathrm{Fe}  & 0& -\Delta_{s-p,d} & \delta_\mathrm{Fe} \\
0& \delta_1 & \delta_\mathrm{Fe} & \Delta_{s-p,d}
\end{bmatrix}
+\HH_{B}, \label{eq:hamiltonian}
\end{equation}
where $\Delta_{s-p,d}$ describes ion-exciton exchange interaction, $\delta_1$ is the effective anisotropic interaction between carriers in the QD and $\delta_\mathrm{Fe}\approx 0$ is an effective mixing energy between spin states of the iron ion due to the crystal environment. The $\HH_B$ term is responsible for additional Zeeman splitting of bright exciton and Fe$^{2+}$ spin in a magnetic field and takes the standard structure $\HH_B={-\mu_B g S_z B}$, where g is the Land$\acute{e}$ factor of either exiton or ion dopant. All parameters used in the hamiltonian \ref{eq:hamiltonian} can be extracted from polarisation-resolved magnetospectroscopic measurements \cite{Smolenski2016}. The same approach also applies to other systems of QD containig TM ions, apart from differences in the ion spin configuration \cite{besombes}.

Invoked Hamiltonian can be used to determine the energy of the eigenstates $\ket{\Psi_i}$ and, together with the dipole moment operators, the oscillator strengths of the optical transitions. However, intensity of photoluminescence lines crucially depends also on the population of the initial states. In order to evaluate these populations we introduce two simplifying assumptions. Firstly, we neglect small changes in the dopant spin upon the QD excitation \cite{pilat1}. Under such assumption, the same probabilities (denoted as $b_1$ and $b_2$) describe the occurrence of up- and down- orientation of the Fe$^{2+}$ spin both in the empty-dot state as in the presence of the exciton. Formally, the effect of the Fe$^{2+}$ spin polarization is accounted for by introducing following factor to population of the $\ket{\Psi_i}$ state:
\begin{displaymath}
\sum_{j=\{1,2\}}b_j|\bra{\mathrm{Fe}_j}P_{\sigma^\pm}\ket{\Psi_i}|^2.
\end{displaymath}

The second assumption is related to the exciton part of the wavefunction, in particular to the degree upon which the circularly polarized excitation leads to formation of the spin-polarized excitons. Namely, we assume that this degree does not depend on the dopant spin and can be described by a single probability $a^+$ that the exciton will be created with spin-up and complementary probability $a^-=1-a^+$ that the exciton will be created with spin-down orientation. Such a description is universal in a sense that it covers a case of quasi-resonant excitation with either $\sigma^+$ or $\sigma^-$ polarization ($a^+ \gg a^-$ and $a^+ \ll a^-$, respectively) or non-resonant excitation case ($a^+ \approx a^-$).

Altogether the discussed assumptions lead to following expression for occupation of the $\ket{\Psi_i}$ state:
\begin{equation}
p_i = \sum_{j=\{1,2\}}b_j\sum_{\sigma^{\pm}}a^{\pm}|\bra{\Psi_i}P_{\sigma^{\pm}}\ket{\mathrm{Fe}_j}|^2.
\end{equation}
The intensity of a particular optical transition is determined by a product of the initial state occupation and the oscillator strength, since for each bright excitonic state the sum of oscillator strengths (and thus the lifetime) is constant:
\begin{equation}
I\left(\ket{\Psi_i}\xrightarrow{\sigma^\pm}\ket{\mathrm{Fe}_j}\right) = p_i\, |\bra{\Psi_i}P_{\sigma^{\pm}}\ket{\mathrm{Fe}_j}|^2 .
\label{eq:intensywnosc}
\end{equation}

Derived expression can be further modified to account for secondary effects, such as partial relaxation of the exciton spin during its lifetime \cite{smolen2015, igor_2015} or probability to flip the dopant spin depending on the orientation of the captured exciton \cite{smolen2018}. However, such corrections are strongly model-dependent and in general require introduction of new free parameters. Thus, for the sake of simplicity, we will remain with pristine version of our model.

Expression \ref{eq:intensywnosc} describes intensity of each of the lines using three parameters (apart from otherwise known parameters of the Hamiltonian): polarization of the Fe$^{2+}$ spin $\mean{S_\mathrm{Fe}}=2\frac{b_1-b_2}{b_1+b_2}$, degree of spin selectivity of the excitons $\mean{\xi_X}=\frac{a^+-a^-}{a^++a^-}$ and the overall intensity of the QD luminescence. Consequently, by means of numerical fitting of experimentally determined intensities of the PL lines in both polarizations of detection, we can extract parameters $a^\pm$ and $b_j$. Such a procedure is general and would work also if the model was additionally extended, as discussed above. However, in the presented form the model can be solved analytically leading to more practical formulas. First of all, by summing intensity detected in both polarizations we obtain particularly simple expression for $\mean{S_\mathrm{Fe}}$:
\begin{equation}
\mean{S_\mathrm{Fe}} = 2\frac{I^\mathrm{(tot)}_{S=+2}-I^\mathrm{(tot)}_{S=-2}}{I^\mathrm{(tot)}_{S=+2}+I^\mathrm{(tot)}_{S=-2}}.
\label{eq:sumowanie}
\end{equation}
It is closely resembling Eq. \ref{eq:stopien}, but the fundamental difference is that $I^\mathrm{(tot)}_{S=\pm 2}$ is a~total intensity of emission lines related to $S=\pm 2$ states \emph{summed over both detection polarizations} under the same excitation conditions. Knowing $b_1 =\frac{1}{2} + \mean{S_\mathrm{Fe}}/4$ and $b_2 =\frac{1}{2} - \mean{S_\mathrm{Fe}}/4$ we can determine degree of spin selectivity of the excitons $\mean{\xi_X}$ using expressions for least squares method:
\begin{equation}
\xi_X = \frac{D(B-C) - E(C-A)}{D(B+C) - E(C+A)}, \label{eq:simplea}
\end{equation}
where $A, \ldots, E$ are defined in Appendix \ref{appendix2}. In principle, quantity $\mean{\xi_X}$ can be also estimated based on intensities of any 2 lines of different polarizations, but utilization of all measured intensities in least squares method leads to more robust estimation.

\section{Application of the model to the interpretation of magneto-optical data}

\begin{figure}[H]
\includegraphics[scale=1]{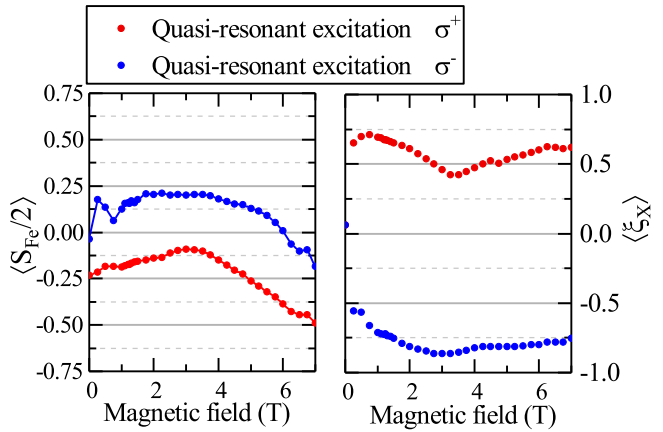}
\caption{\textit{Spin orientations of exciton (a) and Fe$^{2+}$ ion (b) under quasi-resonant excitation as a funciton of magnetic field, determined using proposed model.}}
\label{fig3}
\end{figure}

Figure \ref{fig3} presents the results of applying the described procedure to experimental data from Fig.  \ref{fig2}. The extracted parameters exhibit much weaker variation on the magnetic field than the basic estimator $\eta_\mathrm{Fe}$ discussed earlier. The constructed model unequivocally confirms that most of the disparity between the experimental points in \ref{fig2} originates from the variation of the eigenstates $\ket{\Psi_i}$ rather than from variation of population factors $b_j$ and $a^\pm$. In fact, the latter of the parameters (and consequently also $\mean{\xi_X}$) is nearly independent on the magnetic field. As expected, the principal factor determining the population of the excitonic state is the helicity of the excitation, as manifested by the difference between red and blue series in Fig. \ref{fig3}(b). Similarly, we find clear difference between the average orientation of the iron spin depending on the helicity of the excitation, i.e., the optical orientation effect. As seen in Fig. \ref{fig3}(a) the efficiency of this effect is practically independent on the magnetic field. Given that half of the excitonic states around $B=3$ T are inherently linearly polarized, it might suggest that the leading orientation mechanism occurs at the exciton formation stage rather than during its lifetime. This observation could be compared with the data reported in Ref. \onlinecite{Smolenski2016}, where no decrease in the optical orientation efficiency was observed up to 4 T. On the other hand, the data in Ref. \onlinecite{Smolenski2016} indicated a trend of increasing the efficiency of the optical orientation upon increasing the magnetic field, which is not corroborated here. We tentatively attribute this difference to differences in the energy excitation (2.41 eV compared to 2.54 eV in Ref. \onlinecite{Smolenski2016}), which may affect the relaxation path of the photo-created excitons.

\section{Discussion}
Apart from anisotropic electron-hole exchange interaction analysed in this paper, there are several other physical effects which could affect spin readout and  optical spin orientation in a QD with a single TM ion, such as spin-dependent quenching of excitonic transitions via intra-ionic transitions \cite{Nawrocki_PRB1994,kacper} or relaxation of exciton-magnetic-ion system\cite{Kolpotowski_PRB2011}. Such effects pose serious challenge as they are related to the very idea of the readout of the dopant spin using a~confined exciton and they cannot be resolved without detailed insight into the particular mechanism.

Procedure described in this work addresses a problem of extracting of dopant polarization degree in a case when QD emission lines are not circularly polarized. As we shown, in case of Fe$^{2+}$ ion the effect might even lead to apparent reversing the sign of optical orientation. In principle, effect of the same nature occurs in all systems of QDs with a single TM ions. Yet, it was not recognized in earlier studies as in most cases it does not stand out for one of two reasons. First, typically TM-doped QD exhibit multiple emission lines in each polarization, e.g., 6 lines for Mn-doped CdTe or CdSe QDs. The anticrossings leading to linearly polarized eigenstates are occurring for different pairs of lines at different magnetic fields. As the result, even if a pair emission lines undergoes an anticrossing, the other 4 lines are polarized circularly, and the resulting estimate of averaged dopant spin is not strongly affected. Second, the discussed effect requires efficient transfer of the polarization of the excitation laser to the spin of injected excitons. Under non-resonant excitation the hot excitons do not transfer the polarization, which would correspond to $\xi_X = 0$ in the proposed model. Under such conditions the spin of the dopant remains the only factor influencing the relative populations of the QD states.

\begin{figure}[H]
\centering
\includegraphics{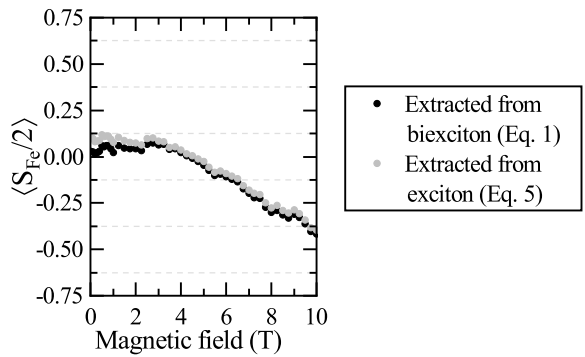}
\caption{
Comparison of Fe$^{2+}$ ion orientation degree under non-resonant excitation derived from biexciton emission intensity and determined using proposed model.} \label{fig4}
\end{figure}

The described procedure enables us to extract the actual polarization of the ion even if the population of the exciton states are perturbed by unequal exciton pumping. Alternative way to address this issue is to use neutral biexciton line, if only it is experimentally viable. As noted earlier, the biexciton recombination is desirable tool for readout of the dopant spin since it is not affected by the spin relaxation due to its singlet nature. For the same reason the relative intensities of the biexciton lines are not affected by the polarization of the excitation, hence in case of the biexciton expression \ref{eq:stopien} is supposed to yield correct results. In order to provide a direct comparison between the results of the proposed method of determination of the ion spin polarization using a neutral exciton and more traditional approach using the neutral biexciton we performed an auxiliary experiment with non-resonant excitation at 3.06 eV. Under such conditions, the neutral biexciton line was clearly visible, which allowed us to cross-check the value of $\mean{S_\mathrm{Fe}}$ extracted using both methods. Although under such high-energy excitation the optical orientation for both exciton and the ion is negligibly small, the comparison between the two methods is an important proof of consistency. Indeed, the experimental data presented in Fig. \ref{fig4} prove that the proposed algorithm yields the same results as the formula \ref{eq:stopien} applied for the neutral biexciton.

\section{Conclusion}
We have presented a method of extracting the exciton and magnetic ion spin orientation in quantum dots with important anisotropy. This approach allowed us to analyze the data under quasi-resonant excitation, with the efficiency of the optical orientation of magnetic ion spin close to 15\%. Our method can be easily expanded to other systems of TM dopants inside semiconductor quantum dots.

\subsection{Acknowledgements}
This work was supported by the Polish Ministry of Science and Higher Education under "Iuventus Plus" project IP2015 031674. This work was also supported by the Polish National Science Centre under decisions DEC-2015/18/E/ST3/00559 and DEC-2016/23/B/ST3/03437. The project was carried out with the use of CePT, CeZaMat, and NLTK infrastructures financed by the European Union - the European Regional Development Fund within the Operational Programme "Innovative economy".

\begin{appendices}

\section{Expression for evaluating neutral exciton spin selectivity \label{appendix2}}
Here we present explicit form of parameters used in Eq. \ref{eq:simplea}, which correspond to applying the least-squares method to equations \ref{eq:intensywnosc} for experimentally obtained
intensities of each transition $I\left(\ket{\Psi_i}\xrightarrow{\alpha}\ket{\mathrm{Fe}_j}\right)$:
\begin{widetext}
\begin{eqnarray}
A & = & \sum_j \sum_i \sum_{\alpha\in\{\sigma^+,\sigma^-\} }  \left( b_j |\bra{\Psi_i}P_{\sigma^+}\ket{\mathrm{Fe}_j}|^2 |\bra{\Psi_i}P_{\alpha}\ket{\mathrm{Fe}_j}|^2 \right)^2, \\
B & = & \sum_j \sum_i \sum_{\alpha\in\{\sigma^+,\sigma^-\} }  \left( b_j |\bra{\Psi_i}P_{\sigma^-}\ket{\mathrm{Fe}_j}|^2 |\bra{\Psi_i}P_{\alpha}\ket{\mathrm{Fe}_j}|^2 \right)^2, \\
C & = & \sum_j \sum_i \sum_{\alpha\in\{\sigma^+,\sigma^-\} }  \left( b_j |\bra{\Psi_i}P_{\alpha}\ket{\mathrm{Fe}_j}|^2 \right)^2 |\bra{\Psi_i}P_{\sigma^+}\ket{\mathrm{Fe}_j}|^2 |\bra{\Psi_i}P_{\sigma^-}\ket{\mathrm{Fe}_j}|^2, \\
D & = & \sum_j \sum_i \sum_{\alpha\in\{\sigma^+,\sigma^-\} }  I\left(\ket{\Psi_i}\xrightarrow{\alpha}\ket{\mathrm{Fe}_j}\right) b_j |\bra{\Psi_i}P_{\sigma^+}\ket{\mathrm{Fe}_j}|^2 |\bra{\Psi_i}P_{\alpha}\ket{\mathrm{Fe}_j}|^2, \\
E & = & \sum_j \sum_i \sum_{\alpha\in\{\sigma^+,\sigma^-\} }  I\left(\ket{\Psi_i}\xrightarrow{\alpha}\ket{\mathrm{Fe}_j}\right) b_j |\bra{\Psi_i}P_{\sigma^-}\ket{\mathrm{Fe}_j}|^2 |\bra{\Psi_i}P_{\alpha}\ket{\mathrm{Fe}_j}|^2 .
\end{eqnarray}
\end{widetext}

\end{appendices}

\end{document}